\documentclass[12pt, draftclsnofoot, onecolumn]{IEEEtran}
\usepackage{graphicx}
\usepackage{amsmath}
\usepackage{amssymb}
\usepackage{bm}
\usepackage[backend=biber, style=ieee, sorting=none]{biblatex}
\usepackage{amsthm}
\usepackage{array}
\theoremstyle{definition}
\newtheorem{lemma}{Lemma}
\newtheorem{proposition}{Proposition}
\usepackage{dsfont}

\usepackage[acronym,nonumberlist,nopostdot,nogroupskip,nomain]{glossaries}
\setacronymstyle{long-short}
\makenoidxglossaries

\newacronym{oodn}{OODN}{on--off digital noise}
\newacronym{bep}{BEP}{bit error probability}
\newacronym{iot}{IoT}{Internet of Things}
\newacronym{awgn}{AWGN}{additive white Gaussian noise}
\newacronym{inr}{INR}{interference-to-noise ratio}
\newacronym{snr}{SNR}{signal-to-noise ratio}
\newacronym{clt}{CLT}{central limit theorem}
\newacronym{lrt}{LRT}{likelihood-ratio test}
\newacronym{cdf}{CDF}{cumulative distribution function}
\newacronym{mc}{MC}{Monte Carlo}
\newacronym{bpsk}{BPSK}{binary phase-shift keying}
\newacronym{bsc}{BSC}{binary symmetric channel}
\newacronym{css}{CSS}{chirp spread spectrum}
\newacronym{nbiot}{NB-IoT}{narrowband Internet of Things}
\newacronym{iid}{i.i.d.}{independent and identically distributed}

\newif\ifappendices
\appendicestrue

\begin{document}

\title{On--Off Digital Noise Modulation under Multi-User Co-Channel Interference}

\author{%
    Daniel C.~Ara\'{u}jo,
    Andr\'{e} A.~dos~Anjos,
    and Hugerles S.~Silva%
    \thanks{D. C. Araújo is with the Telecommunications Laboratory (LabTelecom), University of Brasília (UnB), Brasília, DF 72.444-240, Brazil (e-mail: daniel.araujo@unb.br).}%
    \thanks{A. A. dos Anjos is with the Department of Telecommunications, Federal University of Uberlândia, Patos de Minas 38702-178, Brazil (e-mail: andre.anjos@ufu.br).}%
    \thanks{H. S. Silva is with the Electrical Engineering Department, University of Brasília (UnB), Brasília, DF 70910-900, Brazil (e-mail: hugerles.silva@unb.br).}%
}

\maketitle

\begin{abstract}
This letter analyzes the performance of on-off digital noise (OODN) modulation under multi-user scenarios. While prior works have addressed single-link operation, the impact of co-channel interference remains unexplored. We consider $K$ synchronous OODN interferers over \gls{awgn} and fading channels and derive a unified analytical framework for the bit error probability (BEP). The optimal likelihood-ratio detection threshold is obtained, along with a closed-form expression for the resulting irreducible error floor, which climbs geometrically with the number of co-channel interferers and yields a simple admission-control rule on network density. The analysis is extended to $\kappa$–$\mu$ fading, covering practical millimeter-wave channels with dominant line-of-sight components. Results, corroborated by Monte Carlo simulation, show that the floor is governed by the on-off interference and the non-coherent detector rather than by the noise waveform.
\end{abstract}

\glsresetall


\section{Introduction}
\label{sec:intro}

\IEEEPARstart{M}{assive} \gls{iot} deployments must support many low-complexity devices under stringent energy, hardware, and spectral-efficiency constraints while remaining reliable under interference-limited environments~\cite{Chen}. Noise-based modulation has emerged as a promising candidate in this setting because its inherently non-coherent operation requires no carrier-phase or frequency synchronization~\cite{dasilva}.

Over the past decade, noise-based communication has matured from early environmental- and Johnson-noise concepts, originally motivated by secure, ultra-low-power, and potentially stealth transmission~\cite{Kish_2005,Kish_2006}, into practical modulated-Johnson-noise implementations~\cite{Kapetanovic_2022,Garman_2023} and thermal-noise communication schemes~\cite{Basar2023TCOM}. This line of work culminated in the digital noise-modulation framework of~\cite{Basar2024}, which encodes information directly in the statistics of a random waveform. More recent efforts have broadened the field toward channel estimation from thermal-noise observations~\cite{Shen2025}, noise-domain non-orthogonal multiple access~\cite{Yapici_2024}, security- and reliability-oriented noise-loop modulation~\cite{Mucchi2022}, spread-spectrum and multidimensional noise signaling~\cite{NoiseModSpread,NoiseMod3d}, and differential correlation-based detection~\cite{dbn_2026}. Together, these advances establish noise-based modulation as a rapidly expanding direction for low-complexity non-coherent wireless systems.

Among these approaches, \gls{oodn} modulation conveys information through the presence or absence of a random noise waveform rather than a deterministic carrier~\cite{dosAnjos2025}. Encoding the bit in the signal variance lets OODN combine non-coherent reception with low implementation complexity and robustness to carrier-phase and frequency uncertainties, making it appealing for dense \gls{iot} and machine-type communication~\cite{dosAnjos2025}. Existing studies, however, have focused on point-to-point links over \gls{awgn} and fading channels~\cite{dosAnjos2025,dosanjos2}, whereas practical deployments are multi-user, with terminals sharing spectral resources, and the impact of co-channel transmissions on detection remains unexplored.

In this letter, we investigate for the first time the performance of OODN over \gls{awgn} and fading channels in the presence of $K$ synchronous co-channel OODN interferers. We: (i) express the multi-user \gls{bep} as a weighted combination of interferer-free OODN links; (ii) characterize the likelihood-ratio-optimal threshold for the $K$-user mixture, bracketing it between the known single-link optimum~\cite{dosanjos2} and its all-active value and showing it collapses to the former at high \gls{inr}; (iii) derive a closed-form irreducible interference floor together with the admission-control rule it places on network density; and (iv) extend the analysis to $\kappa$--$\mu$ fading, covering practical indoor line-of-sight and non-line-of-sight $60$\,GHz millimeter-wave conditions.

The remainder of this work is organized as follows. Section~\ref{sec:model} describes the system model and Section~\ref{sec:bep} presents the performance analysis. Section~\ref{sec:results} reports the numerical results and discussion. Finally, Section~\ref{sec:conclusion} concludes.

\section{Multi-User System Model}
\label{sec:model}

\begin{figure}[!t]
\centering
\includegraphics[width=\columnwidth]{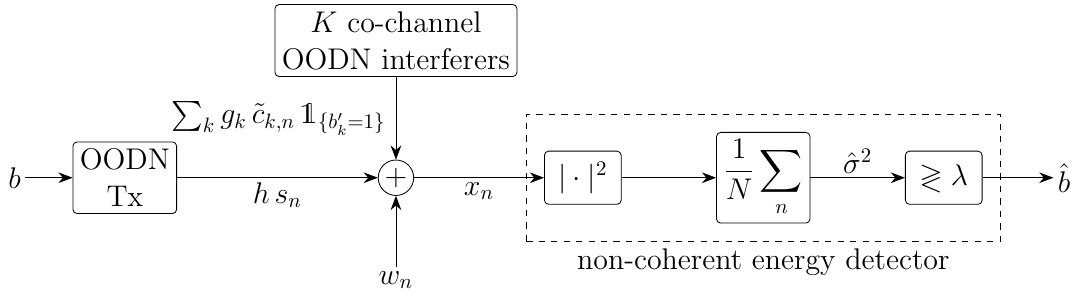}
\caption{Multi-user OODN system model with $K$ co-channel OODN interferers and a non-coherent energy-detection receiver.}
\label{fig:sysmodel}
\end{figure}

We consider a point-to-point OODN link, as depicted in Fig.~\ref{fig:sysmodel}, sharing its band with $K$ co-channel OODN interferers. The desired transmitter encodes each equiprobable bit $b\in\{0,1\}$ into a block of $N$ complex baseband samples as
\begin{equation}
    s_n =
    \begin{cases}
        0, & b=0,\\
        c_n,\ c_n\sim\mathcal{CN}(0,\sigma_{b}^{2}), & b=1,
    \end{cases}
    \qquad n=1,\dots,N,
    \label{eq:signal}
\end{equation}
with $c_n$ \gls{iid} in $n$. Note that interferer $k\in\{1,\dots,K\}$ is an independent OODN node that transmits its own equiprobable bit $b'_k\in\{0,1\}$ over the same interval. It emits $\tilde{c}_{k,n}\sim\mathcal{CN}(0,\sigma_{i}^{2})$ when $b'_k=1$ and stays silent otherwise. At the desired receiver, the $n$-th received sample is given by
\begin{equation}
    x_n \;=\; h\,s_n \;+\; \sum_{k=1}^{K} g_k\,\tilde{c}_{k,n}\,\mathds{1}_{\{b'_k=1\}} \;+\; w_n,
    \label{eq:rx}
\end{equation}
where $w_n\sim\mathcal{CN}(0,\sigma_{w}^{2})$ are \gls{iid} \gls{awgn} samples. The gains $h$ and $\{g_k\}$ are mutually independent block-constant fading gains of the desired and interferer links, respectively. The indicator $\mathds{1}_{\{\cdot\}}$ equals $1$ when its argument holds and $0$ otherwise, so the $k$-th interferer contributes $g_k\tilde{c}_{k,n}$ only while its bit $b'_k=1$. We define the per-sample active-state \gls{snr} $\gamma_S\triangleq|h|^{2}\sigma_{b}^{2}/\sigma_{w}^{2}$, the active-state \gls{inr} of the $k$-th interferer $\gamma_{I,k}\triangleq|g_k|^{2}\sigma_{i}^{2}/\sigma_{w}^{2}$, and the normalized detection threshold $\tau\triangleq\lambda/\sigma_{w}^{2}$, where $\lambda$ is the decision threshold applied to the empirical-variance statistic defined below.
We treat the symmetric homogeneous network in which all interferers share a common \gls{inr} $\gamma_I\triangleq\gamma_{I,k}$. The OODN receiver is non-coherent: it forms the empirical variance
\begin{equation}
    \hat{\sigma}^{2} \;=\; \frac{1}{N}\sum_{n=1}^{N}|x_n|^{2}
    \label{eq:estimator}
\end{equation}
and decides $\hat{b}=1$ if $\hat{\sigma}^{2}\geq\lambda$ and $\hat{b}=0$ otherwise. The interferers are assumed bit-synchronous with the desired link.

\section{Performance Analysis}
\label{sec:bep}

\subsection{Conditional Distribution of the Decision Statistic}
\label{ssec:dist}

Let $\bm{b}'=(b'_1,\dots,b'_K)\in\{0,1\}^{K}$ be the interferer activity pattern and let $m\triangleq\sum_{k=1}^{K} b'_k$ be the number of active interferers. Conditioned on $(b,\bm{b}',h,\{g_k\})$, the three terms in~\eqref{eq:rx} are mutually independent zero-mean complex Gaussians, so $x_n\sim\mathcal{CN}(0,\sigma_{v}^{2})$ is \gls{iid} with
\begin{equation}
    \frac{\sigma_{v}^{2}}{\sigma_{w}^{2}} \;=\; 1 + \gamma_S\,\mathds{1}_{\{b=1\}} + m\,\gamma_I.
    \label{eq:sigmav}
\end{equation}
Crucially, the interferers contribute no mean shift, only an increase of the conditional variance proportional to the number of active nodes $m$. The variance estimator presented in~\eqref{eq:estimator} is therefore a sum of $N$ \gls{iid} central chi-square contributions, $2N\hat{\sigma}^{2}/\sigma_{v}^{2}\sim\chi^{2}_{2N}$, whose \gls{cdf}, for any integer $N\geq1$, is written in terms of the elementary Erlang form as
\begin{equation}
    F_{2N}(y) \;\triangleq\; \Pr[\chi^{2}_{2N}<y] \;=\; 1 - e^{-y/2}\sum_{k=0}^{N-1}\frac{(y/2)^{k}}{k!}.
    \label{eq:erlang}
\end{equation}
Because $b'_1,\dots,b'_K$ are \gls{iid} and equiprobable, the active count is binomial, $m\sim\mathrm{Bin}(K,\tfrac{1}{2})$, so $\Pr[m=j]=\binom{K}{j}2^{-K}$.

\subsection{Exact BEP}
\label{ssec:bep}

\begin{lemma}
\label{lem:mixture}
Conditioned on $(h,\{g_k\})$, the \gls{bep} of OODN under $K$ synchronous and equal-\gls{inr} co-channel OODN interferers is
\begin{equation}
    P_b^{(K)}(\tau) \;=\; 2^{-K}\sum_{m=0}^{K}\binom{K}{m}\,
    P_b^{(0)}\!\Bigl(\tfrac{\gamma_S}{1+m\gamma_I},\,\tfrac{\tau}{1+m\gamma_I}\Bigr),
    \label{eq:bep_mu_struct}
\end{equation}
where
\begin{equation}
    P_b^{(0)}(\gamma_S,\tau) \triangleq \tfrac{1}{2}\!\left[1 - F_{2N}(2N\tau) + F_{2N}\!\bigl(\tfrac{2N\tau}{1+\gamma_S}\bigr)\right]
    \label{eq:baseline}
\end{equation}
is the interferer-free OODN \gls{bep}, discussed in~\cite{dosAnjos2025}, evaluated at \gls{snr} $\gamma_S$ and normalized threshold $\tau$.
\end{lemma}

\ifappendices Proof: See Appendix~\ref{app:mixture}. \qed \fi

Equation~\eqref{eq:bep_mu_struct} admits an operational reading. A $K$-user OODN network behaves as a $2^{-K}$-weighted mixture of $2^{K}$ interferer-free OODN links. Here $m$ active neighbors divide both the effective \gls{snr} $\gamma_S$ and the threshold $\tau$ by the same factor $1+m\gamma_I$. Substituting the Erlang form~\eqref{eq:erlang} into each mixture term, the \gls{bep} reduces to a finite sum of elementary exponentials, an expression that involves neither Marcum-$Q$ nor incomplete-gamma functions. For $K=1$ it reduces to the single-interferer case, and for $\gamma_I\to0$ it collapses to the interferer-free baseline $P_b^{(0)}(\gamma_S,\tau)$. The heterogeneous case, in which interferers carry distinct \glspl{inr} $\gamma_{I,k}$, is recovered by replacing the binomial mixture in~\eqref{eq:bep_mu_struct} with a sum over the $2^{K}$ equiprobable activity vectors $\bm{b}'$, yielding
\begin{equation}
    P_b^{(K)}(\tau) \;=\; 2^{-K}\!\!\sum_{\bm{b}'\in\{0,1\}^{K}}\!\!
    P_b^{(0)}\!\Bigl(\tfrac{\gamma_S}{1+\Gamma(\bm{b}')},\,\tfrac{\tau}{1+\Gamma(\bm{b}')}\Bigr),
    \label{eq:bep_hetero}
\end{equation}
where $\Gamma(\bm{b}')\triangleq\sum_{k=1}^{K}b'_k\,\gamma_{I,k}$ is the aggregate \gls{inr} of the active interferers. When all $\gamma_{I,k}=\gamma_I$, $\Gamma(\bm{b}')$ depends only on the active count $m$ and~\eqref{eq:bep_hetero} reduces to the binomial mixture in~\eqref{eq:bep_mu_struct}.

\subsection{Likelihood-Ratio-Optimal Threshold}
\label{ssec:threshold}

Because the conditional law of $\hat{\sigma}^{2}$ involves only central chi-square variates, the Bayes-optimal detector is a \gls{lrt}. For the interferer-free OODN link, the optimal threshold was derived in closed form in~\cite{dosanjos2} as
\begin{equation}
    \tau^{\star}_{(0)}(\gamma_S) \;=\; \frac{(1+\gamma_S)\ln(1+\gamma_S)}{\gamma_S},
    \label{eq:tau_lrt}
\end{equation}
which is exact and independent of $N$. We extend this single-link result to the $K$-user mixture. Applying~\eqref{eq:tau_lrt} to each term of~\eqref{eq:bep_mu_struct} gives the per-term optima $\tau_m^{\star}=(1+m\gamma_I)\,\tau^{\star}_{(0)}(\gamma_S/(1+m\gamma_I))$, which are increasing in $m$. The Bayes-optimal threshold of the full mixture satisfies the \gls{lrt} condition
\begin{equation}
    \sum_{m=0}^{K}\binom{K}{m}\!\left[\frac{e^{-N\tau/s_{1,m}}}{s_{1,m}^{N}} - \frac{e^{-N\tau/s_{0,m}}}{s_{0,m}^{N}}\right] = 0,
    \label{eq:lrt_mixture}
\end{equation}
with $s_{0,m}=1+m\gamma_I$ and $s_{1,m}=1+\gamma_S+m\gamma_I$. Each mixture term contributes an exponential in $\tau$ whose decay rate, $N/s_{0,m}$ or $N/s_{1,m}$, depends on $m$. Because these rates differ from term to term, the sum cannot be collapsed into a single exponential, so, unlike the two-term baseline~\eqref{eq:tau_lrt}, the mixture optimum has no closed form. It is nonetheless easy to locate. The per-term optima $\tau_m^{\star}$ increase with $m$, so the mixture optimum lies between the silent-term value $\tau_0^{\star}=\tau^{\star}_{(0)}(\gamma_S)$ and the all-active value $\tau_K^{\star}$. The root of~\eqref{eq:lrt_mixture} is thus bracketed by $[\tau_0^{\star},\tau_K^{\star}]$ and found in a few Newton or bisection steps. At high \gls{inr} the silent term dominates and the optimum collapses to the baseline $\tau^{\star}_{(0)}(\gamma_S)$ of~\eqref{eq:tau_lrt}, which we therefore adopt as the operating threshold in Section~\ref{sec:results}.

\subsection{Irreducible Interference Floor}
\label{ssec:floor}


\begin{proposition}
\label{prop:floor}
For every $\gamma_S\geq0$, $N\geq1$, and fixed threshold $\tau>0$, the \gls{bep} converges, in the interference-limited regime, to the irreducible error floor $\Phi_K$ given by
\begin{equation}
    \Phi_K \;\triangleq\; \lim_{\gamma_I\to\infty} P_b^{(K)}(\tau)
    \;=\; \tfrac{1}{2}\bigl(1-2^{-K}\bigr) \;+\; 2^{-K}\,P_b^{(0)}(\gamma_S,\tau).
    \label{eq:floor}
\end{equation}
As $\gamma_S\to\infty$, $\Phi_K\to\tfrac{1}{2}(1-2^{-K})$.
\end{proposition}

\ifappendices Proof: See Appendix~\ref{app:floor}. \qed \fi

Proposition~\ref{prop:floor}, our central result, shows that the floor is dictated by the on--off structure of the interference rather than by the noise waveform. The limit defining $\Phi_K$ is the high-\gls{inr} asymptote of the interference-limited regime rather than a literal infinite-power assumption, since the \gls{bep} saturates at $\Phi_K$ as soon as an active co-channel neighbor dominates the noise. The desired bit is recoverable only in the all-silent event, of probability $2^{-K}$, in which no neighbor transmits. In each of the remaining $2^{K}-1$ events at least one active interferer inflates the variance estimator and forces the receiver to decide bit~$1$, so the conditional \gls{bep} is exactly $\tfrac12$. Averaging these two outcomes leaves the error-free fraction $2^{-K}$ shrinking geometrically in $K$, and at high desired \gls{snr} the residual term $2^{-K}P_b^{(0)}$ vanishes, so the floor reduces to $\tfrac12(1-2^{-K})$ and takes the values $\Phi_1=\tfrac14$, $\Phi_2=\tfrac38$, $\Phi_3=\tfrac{7}{16}$, approaching $\tfrac12$ as $K\to\infty$. Three consequences follow. First, because the asymptotic high-\gls{snr} floor is independent of $\gamma_S$ and $N$, neither extra power nor a longer observation window on the desired link can break the floor once a neighbor is active. The floor can be reduced only by limiting the number of active neighbors. Second, a single co-channel OODN neighbor reaches a floor of $1/4$, markedly below the $1/2$ floor of a constant-modulus interferer, but this advantage erodes geometrically as the network densifies, with $\Phi_K\to1/2$. Third, inverting the closed-form floor $\tfrac12(1-2^{-K})$ exposes a hard feasibility limit: meeting a target $\Phi^{\star}$ requires $K\leq\log_2\!\bigl(\tfrac{1}{1-2\Phi^{\star}}\bigr)$, so any useful target $\Phi^{\star}<\tfrac14$ forces $K=0$. The attainable range $[\tfrac14,\tfrac12)$ is unusable for an uncoded link, so synchronous co-channel reuse of OODN requires an orthogonal medium-access layer.

\subsection{Extension to $\kappa$--$\mu$ Fading}
\label{ssec:fading}

We now let the desired and interferer links undergo $\kappa$--$\mu$ fading~\cite{Yacoub2007}, a general line-of-sight model whose unit-mean power density is
\begin{equation}
    f(\omega) = \frac{\mu(1+\kappa)^{\frac{\mu+1}{2}}}{\kappa^{\frac{\mu-1}{2}}e^{\mu\kappa}}\,\omega^{\frac{\mu-1}{2}}\,e^{-\mu(1+\kappa)\omega}\,I_{\mu-1}\!\bigl(2\mu\sqrt{\kappa(1+\kappa)\omega}\bigr),
    \label{eq:kappamu_pdf}
\end{equation}
where $I_{\nu}(\cdot)$ is the modified Bessel function, $\kappa\geq0$ is the dominant-to-scattered power ratio, and $\mu>0$ relates to the number of multipath clusters. We write $X\sim\kappa$--$\mu(\kappa,\mu)$ to mean that the unit-mean power variate $X$ has the density~\eqref{eq:kappamu_pdf} with parameters $\kappa$ and $\mu$. The model is general: it includes Rayleigh ($\mu=1,\kappa\to0$), Nakagami-$m$ ($\kappa\to0,\mu=m$), Rice ($\mu=1$), and the one-sided Gaussian as special cases. Writing $\gamma_S=|h|^2\bar{\gamma}_S$ with average \gls{snr} $\bar{\gamma}_S$ and per-user average \gls{inr} $\bar{\gamma}_I$, and since the sum of $m$ \gls{iid} $\kappa$--$\mu$ power variates is again $\kappa$--$\mu$ with parameters $(\kappa,m\mu)$~\cite[Sec.~2.7]{Yacoub2007}, the average \gls{bep} follows by averaging~\eqref{eq:bep_mu_struct} over the channel gains,
\begin{equation}
    \bar{P}_b^{(K)} = 2^{-K}\!\sum_{m=0}^{K}\!\binom{K}{m}
    \mathbb{E}_{\Omega_h,\Omega_m}\!\left[P_b^{(0)}\!\Bigl(\tfrac{\bar{\gamma}_S\Omega_h}{1+m\bar{\gamma}_I\Omega_m},\tfrac{\tau}{1+m\bar{\gamma}_I\Omega_m}\Bigr)\right],
    \label{eq:avg_bep}
\end{equation}
where $\mathbb{E}_{\Omega_h,\Omega_m}[\cdot]$ denotes expectation over the independent normalized powers $\Omega_h$ and $\Omega_m$, with $\Omega_h\sim\kappa$--$\mu(\kappa,\mu)$ the desired power and $\Omega_m\sim\kappa$--$\mu(\kappa,m\mu)$ the aggregate power of the $m$ active interferers. Note that each expectation is a low-dimensional integral evaluated efficiently by Gauss--Laguerre quadrature in analogy with~\cite[Eq.~(16)]{dosAnjos2025}. The interference floor is unaffected: applying~\eqref{eq:floor} pointwise and invoking dominated convergence, since each summand is bounded by $\tfrac12$ and $|h|^2,|g_k|^2>0$ almost surely,
\begin{equation*}
    \lim_{\bar{\gamma}_I\to\infty}\bar{P}_b^{(K)} \;=\; \tfrac{1}{2}\bigl(1-2^{-K}\bigr) + 2^{-K}\,\bar{P}_b^{(0)}(\bar{\gamma}_S),
\end{equation*}
so the $\Phi_K$ structure governs the multi-user OODN link under general $\kappa$--$\mu$ fading exactly as under \gls{awgn}.

\section{Numerical Results}
\label{sec:results}

We validate the closed-form expressions against \gls{mc} simulation for $N=100$ samples per bit. The theoretical curves use the mixture \gls{bep}~\eqref{eq:bep_mu_struct} evaluated at the closed-form baseline \gls{lrt} threshold $\tau^{\star}_{(0)}(\gamma_S)$ of~\eqref{eq:tau_lrt}, the high-\gls{inr} limit of the mixture optimum. Computed from the fixed $\gamma_S$, this threshold is held constant as $\gamma_I$ varies and thus satisfies the fixed-$\tau$ premise of Proposition~\ref{prop:floor}, so each curve saturates at the corresponding floor $\Phi_K$. In all figures, lines show the closed-form theory and markers show \gls{mc} estimates accumulated to at least $2000$ bit errors per point.

\begin{figure}[t]
\centering
\includegraphics{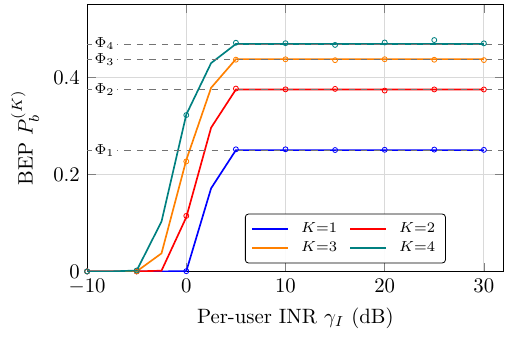}
\caption{BEP versus per-user INR for OODN under $K=1$ to $4$ co-channel OODN interferers ($\gamma_S=10$\,dB, $N=100$).}
\label{fig:floor}
\end{figure}

Fig.~\ref{fig:floor} reports the \gls{bep} as the per-user \gls{inr} is swept at $\gamma_S=10$\,dB and $N=100$, for $K=1$ to $4$ co-channel interferers, with OODN detected at its baseline-optimal threshold. Each curve saturates at exactly the analytic floor $\Phi_K$ of Proposition~\ref{prop:floor}, confirming that the floor is set by the on--off Gaussian multiple-access interference and the energy detector rather than by the noise waveform. The monotonic increase $\Phi_1=\tfrac14\!\to\!\Phi_2=\tfrac38\!\to\!\Phi_3=\tfrac{7}{16}\!\to\!\Phi_4=\tfrac{15}{32}$ quantifies how the robustness of OODN to a single stochastic interferer degrades as the number $K$ of co-channel users grows.

\begin{figure}[t]
\centering
\includegraphics{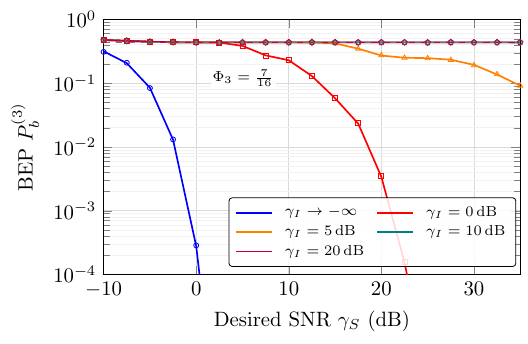}
\caption{BEP of OODN versus desired SNR for $K=3$ at five per-user INR levels ($N=100$).}
\label{fig:waterfall}
\end{figure}

Fig.~\ref{fig:waterfall} complements the \gls{inr} sweeps by plotting the \gls{bep} of OODN against the desired \gls{snr} $\gamma_S$ for $K=3$ at five per-user \gls{inr} levels, $\gamma_I\in\{-\infty,0,5,10,20\}$\,dB. In the interferer-free case, the \gls{bep} decays monotonically toward the estimator-variance floor $Q(\sqrt{N})$, whereas every interfered curve saturates at the analytic floor $\Phi_3=7/16$. Once an active neighbor is present, no increase in $\gamma_S$ drives the \gls{bep} below $\Phi_3$, which confirms that the floor is independent of the desired \gls{snr}. At low \gls{snr}, the horizontal spacing between adjacent curves reproduces the effective-\gls{snr} penalty $1+m\gamma_I$ incurred by the active terms of the mixture.

\begin{figure}[t]
\centering
\includegraphics{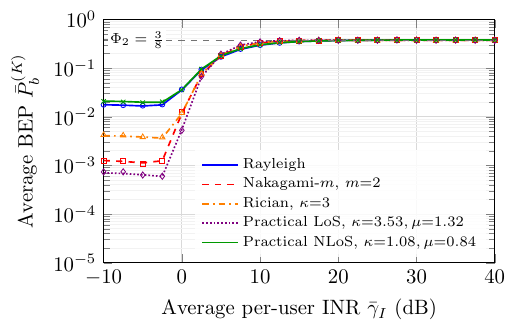}
\caption{Average BEP versus average per-user INR under $\kappa$--$\mu$ fading for $K=2$ ($\bar{\gamma}_S=20$\,dB, $N=100$).}
\label{fig:kappamu}
\end{figure}

Finally, Fig.~\ref{fig:kappamu} reports the average \gls{bep} defined in~\eqref{eq:avg_bep} versus the average per-user \gls{inr} under $\kappa$--$\mu$ fading, for $K=2$ and fixed $\bar{\gamma}_S=20$\,dB. The curves include representative special cases of the $\kappa$--$\mu$ model, namely Rayleigh fading $(\kappa=0,\mu=1)$, Nakagami-$m$ fading $(\kappa=0,\mu=2)$, and Rician fading $(\kappa=3,\mu=1)$, together with two practical indoor 60~GHz channel configurations reported in~\cite{reis_2019}: a LoS case with $\kappa=3.53$, $\mu=1.32$, and Tx--Rx distance of 2.77~m, and an NLoS case with $\kappa=1.08$, $\mu=0.84$, and Tx--Rx distance of 3.29~m. These practical parameters were extracted from measurements conducted in a laboratory indoor environment with a high density of scatterers and reflectors, using omnidirectional antennas at both the transmitter and receiver. At low \gls{inr}, the system is fading-limited, and the performance strongly depends on the channel parameters. The practical LoS channel provides the lowest \gls{bep}, whereas Rayleigh and the practical NLoS channel yield the worst performance. As the average \gls{inr} increases, the curves rapidly lose their dependence on the desired-link fading statistics and approach the common high-\gls{snr} asymptotic floor $\Phi_2=3/8$ predicted by Proposition~\ref{prop:floor}. This confirms that, in the interference-limited regime, the limiting performance is governed primarily by the activity of the co-channel OODN interferers rather than by the small-scale fading distribution of the desired link.

\section{Conclusion}
\label{sec:conclusion}

This letter derived the exact \gls{bep} of OODN under $K$ synchronous co-channel OODN interferers, the dominant impairment in massive multi-user \gls{iot} deployments. Characterizing the decision statistic as a binomial mixture of central chi-square variates yields a closed-form \gls{bep}, a $2^{-K}$-weighted mixture of interferer-free OODN links, together with an \gls{lrt}-optimal mixture threshold that collapses at high \gls{inr} to the known $N$-independent single-link optimum. The irreducible floor equals $\Phi_K=\tfrac12(1-2^{-K})$. Since even a single active neighbor caps the \gls{bep} at $1/4$, synchronous co-channel operation requires an orthogonal medium-access layer. All results were corroborated by \gls{mc} and persist under $\kappa$--$\mu$ fading, including practical channels parameterized from $60$\,GHz indoor measurements. 

\ifappendices
\appendices

\section{Derivation of the Mixture BEP (Lemma~\ref{lem:mixture})}
\label{app:mixture}

Fix the bit $b$, the activity pattern $\bm{b}'$ with $m=\sum_k b'_k$, and the channel gains. By~\eqref{eq:rx}, the samples are \gls{iid}, $x_n\sim\mathcal{CN}(0,\sigma_v^2)$, with conditional variance $\sigma_v^2=\sigma_w^2(1+\gamma_S\mathds{1}_{\{b=1\}}+m\gamma_I)$ from~\eqref{eq:sigmav}. The estimator~\eqref{eq:estimator} therefore obeys $2N\hat{\sigma}^2/\sigma_v^2\sim\chi^2_{2N}$, and the decision rule compares $\hat{\sigma}^2$ with $\lambda=\tau\sigma_w^2$. Since $\sigma_v^2/\sigma_w^2=1+m\gamma_I$ for $b=0$ and $1+\gamma_S+m\gamma_I$ for $b=1$, the false-alarm and miss probabilities follow from~\eqref{eq:erlang},
\begin{align}
    \Pr[\hat{\sigma}^2\!\geq\!\lambda\mid b{=}0,m] &= 1-F_{2N}\!\Bigl(\tfrac{2N\tau}{1+m\gamma_I}\Bigr),
    \label{eq:app_pe0}\\
    \Pr[\hat{\sigma}^2\!<\!\lambda\mid b{=}1,m] &= F_{2N}\!\Bigl(\tfrac{2N\tau}{1+\gamma_S+m\gamma_I}\Bigr).
    \label{eq:app_pe1}
\end{align}

Averaging these with weight $\tfrac12$ and writing $\delta\triangleq1+m\gamma_I$, the error probability given $m$ is
\begin{align}
    P_b(m)
    &= \tfrac12\Bigl[1-F_{2N}\!\bigl(\tfrac{2N\tau}{\delta}\bigr)+F_{2N}\!\bigl(\tfrac{2N\tau}{\delta+\gamma_S}\bigr)\Bigr]
       \notag\\
    &= \tfrac12\Bigl[1-F_{2N}\!\bigl(2N\tfrac{\tau}{\delta}\bigr)
       +F_{2N}\!\bigl(\tfrac{2N(\tau/\delta)}{1+\gamma_S/\delta}\bigr)\Bigr]
       \label{eq:app_pbm2}\\
    &= P_b^{(0)}\!\Bigl(\tfrac{\gamma_S}{\delta},\,\tfrac{\tau}{\delta}\Bigr),
       \label{eq:app_pbm3}
\end{align}
where~\eqref{eq:app_pbm3} applies the definition~\eqref{eq:baseline} at \gls{snr} $\gamma_S/\delta$ and threshold $\tau/\delta$.

Since $m\sim\mathrm{Bin}(K,\tfrac12)$, averaging over $m$ gives
\begin{align*}
    P_b^{(K)}(\tau)
    &= \sum_{m=0}^{K}\Pr[m]\,P_b(m)\\
    &= 2^{-K}\sum_{m=0}^{K}\binom{K}{m}\,P_b^{(0)}\!\Bigl(\tfrac{\gamma_S}{1+m\gamma_I},\tfrac{\tau}{1+m\gamma_I}\Bigr),
\end{align*}
which is~\eqref{eq:bep_mu_struct}. This completes the proof. \qed

\section{Derivation of the Interference Floor (Proposition~\ref{prop:floor})}
\label{app:floor}

We evaluate $\lim_{\gamma_I\to\infty}$ of~\eqref{eq:bep_mu_struct} term by term.

Consider an active term, $m\geq1$, with $\delta\triangleq 1+m\gamma_I$, i.e.\ $P_b^{(0)}(\gamma_S/\delta,\tau/\delta)$. From~\eqref{eq:erlang}, the chi-square \gls{cdf} vanishes at the origin,
\begin{equation}
    F_{2N}(0) = 1 - e^{0}\sum_{k=0}^{N-1}\frac{0^{k}}{k!} = 1 - 1 = 0,
    \label{eq:app_F0}
\end{equation}
and $F_{2N}$ is continuous. As $\gamma_I\to\infty$, $\delta\to\infty$ and both arguments vanish, so by~\eqref{eq:app_F0},
\begin{align*}
    P_b^{(0)}\!\Bigl(\tfrac{\gamma_S}{\delta},\tfrac{\tau}{\delta}\Bigr)
    &= \tfrac12\Bigl[\,1 - F_{2N}\!\bigl(\tfrac{2N\tau}{\delta}\bigr)
       + F_{2N}\!\bigl(\tfrac{2N\tau}{\delta+\gamma_S}\bigr)\Bigr]\\
    &\xrightarrow{\gamma_I\to\infty}\; \tfrac12\,[\,1-0+0\,]\;=\;\tfrac12 .
\end{align*}
Operationally, $\tau/\delta$ falls below the noise level, so the detector declares $\hat{b}=1$ regardless of $b$ and errs on half the bits.

The silent term $m=0$ is independent of $\gamma_I$ and equals $P_b^{(0)}(\gamma_S,\tau)$. Summing both with $\sum_{m=1}^{K}\binom{K}{m}=2^{K}-1$, it follows
\begin{align*}
    \Phi_K &= 2^{-K}\Bigl[P_b^{(0)}(\gamma_S,\tau)+\tfrac12(2^{K}-1)\Bigr]\\
           &= \tfrac12(1-2^{-K})+2^{-K}P_b^{(0)}(\gamma_S,\tau),
\end{align*}
which is~\eqref{eq:floor}. As $\gamma_S\to\infty$ with threshold in $(1,1+\gamma_S)$, both $F_{2N}$ terms of $P_b^{(0)}(\gamma_S,\tau)$ vanish, so $\Phi_K\to\tfrac12(1-2^{-K})$. This completes the proof. \qed
\fi

\printbibliography

\end{document}